\documentclass[english,reqno]{smfart}

\usepackage{epsfig,times}

\usepackage[mathcal]{euscript}
\usepackage{amsmath}
\usepackage{amsthm}
\usepackage{amsfonts}
\usepackage{amssymb}
\newcommand{\fr}[2]{{\textstyle \frac{#1}{#2} }}
\newcommand{\ga}{\gamma}
\newcommand{\ot}{\otimes}

\def\1{\one}
\def\2{\two}

\theoremstyle{plain}
\newtheorem{thm}{Theorem}

\theoremstyle{remark}

\newcommand{\CA}{{\mathcal A}}

\newcommand{\CC}{{\mathcal C}}
\newcommand{\CD}{{\mathcal D}}

\newcommand{\CF}{{\mathcal F}}
\newcommand{\CG}{{\mathcal G}}
\newcommand{\CH}{{\mathcal H}}

\newcommand{\CP}{{\mathcal P}}

\newcommand{\CT}{{\mathcal T}}

\newcommand{\CU}{{\mathcal U}}
\newcommand{\CV}{{\mathcal V}}

\newcommand{\SC}{{\mathsf C}}
\newcommand{\SD}{{\mathsf D}}

\newcommand{\SL}{{\mathsf L}}

\newcommand{\SR}{{\mathsf R}}

\newcommand{\SU}{{\mathsf U}}
\newcommand{\SV}{{\mathsf V}}

\newcommand{\fsl}{{\mathfrak s}{\mathfrak l}}

\newcommand{\sfc}{{\mathsf c}}

\renewcommand{\sf}{{\mathsf f}}
\newcommand{\sh}{{\mathsf h}}
\newcommand{\sll}{{\mathsf l}}

\newcommand{\sq}{{\mathsf q}}
\newcommand{\spp}{{\mathsf p}}

\newcommand{\mss}{{\mathsf s}}

\newcommand{\sx}{{\mathsf x}}

\newcommand{\sz}{{\mathsf z}}

\newcommand{\al}{a} 
\newcommand{\be}{\beta}
\renewcommand{\ga}{\gamma}
\newcommand{\Ga}{\Gamma}
\newcommand{\de}{\delta}
\newcommand{\De}{\Delta}

\newcommand{\one}{{\mathfrak 1}}
\newcommand{\two}{{\mathfrak 2}}

\newcommand{\BR}{{\mathbb R}}

\newcommand{\BI}{{\mathbb I}}
\newcommand{\BC}{{\mathbb C}}
\newcommand{\BP}{{\mathbb P}}
\newcommand{\BS}{{\mathbb S}}

\newcommand{\BZ}{{\mathbb Z}}

\begin{document}                 

\title{On the relation between quantum Liouville
theory and the quantized Teichm\"uller spaces}
\author{J. Teschner}
\address{Institut f\"ur theoretische Physik\\                 
Freie Universit\"at Berlin,\\                        
Arnimallee 14\\                                    
14195 Berlin\\ Germany}

\begin{abstract}We review both the construction of conformal blocks
in quantum Liouville theory and the quantization of 
Teichm\"uller spaces as developed by Kashaev, Checkov and Fock.
In both cases one assigns to a Riemann surface 
a Hilbert space acted on by a representation of the
mapping class group. According to a conjecture of H. Verlinde,
the two are equivalent. We describe some key steps in the
verification of this conjecture.
\end{abstract}

\maketitle

\begin{center}
\medskip
\em{Dedicated to A.A.~Belavin on his $60^{\rm th}$ birthday} 
\medskip
\end{center}

\section{Introduction}

Quantum Liouville theory is a crucial ingredient for a variety 
of models for low dimensional quantum gravity
and non-critical string theories. In the case of two dimensional 
quantum gravity or non-critical string theories this is a consequence 
of the Weyl-anomaly \cite{Pol}, which forces one to 
take into account the quantum dynamics of the conformal factor of the
two-dimensional metric. More recently it was proposed that 
Liouville theory also plays a crucial role for {\it three-dimensional} quantum
gravity in the presence of a cosmological constant in the sense of
representing a holographic dual for this theory, see e.g. \cite{Kr,KV}
and references therein.

For all these applications it is crucial that the quantum Liouville theory
has a geometric interpretation as describing the  quantization of spaces
of two-dimensional metrics. Such an interpretation is to be expected
due to the close connections between {\it classical} Liouville theory and 
the theory of Riemann surfaces. Having fixed a complex structure on the
Riemann surface one may represent the 
unique metric of negative constant curvature locally in the form 
$ds^2=e^{2\varphi}dzd\bar{z}$ where $\varphi$ must solve the
Liouville equation $\partial\bar{\partial}\varphi=\frac{1}{4}e^{2\varphi}$.
This relation between the Liouville equation and the uniformization 
problem leads to beautiful connections between classical Liouville theory
and the theory of moduli spaces of Riemann surfaces \cite{ZT}.

One may therefore expect that quantum Liouville theory is related
to some sort of ``quantization'' of the moduli spaces of  
Riemann surfaces. A concrete proposal in this direction was made by
H. Verlinde in \cite{V}, where it was proposed that the space of 
conformal blocks of the Liouville theory with its mapping class 
group representation is isomorphic to the 
space of states obtained by quantizing the Teichm\"uller spaces of
Riemann surfaces \cite{Fo,Ka1,CF}. We believe that understanding
Verlinde's conjecture will serve as a useful starting point for
developing the geometrical interpretation of the quantized Liouville theory
in general. The aim of the present paper
will be to outline the definition of the objects that are
involved in Verlinde's conjecture and to describe some key steps
towards the proof of it.

\section{Teichm\"uller spaces} 

The Teichm\"uller spaces $\CT(\Sigma)$ are the spaces
of deformations of the complex structures on Riemann surfaces $\Sigma$.
As there is a unique metric of constant curvature -1 associated to 
each complex structure one may identify the Teichm\"uller spaces
with the spaces of deformations of the metrics with constant curvature -1.
Coordinates for the Teichm\"uller spaces can therefore be obtained
by considering the geodesics that are
defined by the constant curvature metrics.

\subsection{Penner coordinates}

\begin{figure}[h]
\begin{center}\epsfxsize7cm
\epsfbox{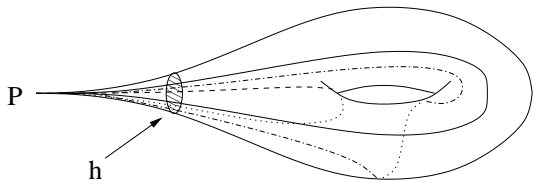}\hspace{1cm}
\end{center}
\caption{Triangulation of the once-punctured torus.}\label{triang}
\end{figure}

A particularly useful set of coordinates was introduced
by R. Penner in \cite{Pe}. They can be defined for Riemann surfaces 
that have at
least one puncture. One may assume having triangulated the surface 
by geodesics that start and end at the punctures. As an example we have 
drawn in Figure \ref{triang} a triangulation of
the once-punctured torus.
The length of these
geodesics will be infinite. In order to regularize this divergence
one may introduce one horocycle around each puncture and measure 
only the length of the segment of a geodesic that lies between the 
horocycles. Assigning to an edge $e$ its regularized length $l_e$
gives coordinates for the so-called decorated Teichm\"uller spaces. 
These are fiber spaces over the
Teichm\"uller spaces which have fibers that parameterize the choices of the 
``cut-offs'' as introduced by the horocycles. 

\begin{figure}[h]
\centering\setlength{\unitlength}{0.02in}
\begin{picture}(200,40)
\put(100,0){\line(-1,1){20}}\put(85,3){$d$}
\put(100,0){\line(1,1){20}}\put(112,3){$c$}
\put(80,20){\line(1,1){20}}\put(85,32){$a$}
\put(120,20){\line(-1,1){20}}\put(112,32){$b$}
\put(100,0){\line(0,1){40}}\put(102,18){$e$}
\put(100,0){\circle*{3}}
\put(80,20){\circle*{3}}
\put(100,40){\circle*{3}}
\put(120,20){\circle*{3}}
\end{picture}
\caption{The labeling of the edges}\label{rectlab}
\end{figure}
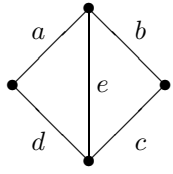

A closely related set of coordinates for 
the Teichm\"uller spaces themselves was introduced by Fock in \cite{Fo}.
The coordinate $z_e$ associated to an edge $e$ of a triangulation
can be expressed in terms of the Penner-coordinates via
$z_e=l_a+l_c-l_b-l_d$, where $a$, $b$, $c$ and $d$ label the
other edges of the triangles that have $e$ in its boundary as indicated in
Figure \ref{rectlab}. 

Instead of triangulations of the Riemann surfaces
it is often convenient to consider the corresponding {\it fat graphs}, which
are defined by putting a trivalent vertex into each triangle and by 
connecting these vertices such that the edges of the triangulation are 
in one-to-one correspondence to the edges of the fat-graph.

\subsection{Symplectic structure}

The Teichm\"uller spaces carry a natural symplectic form, called 
Weil-Petersson symplectic form \cite{IT}. We are therefore dealing with a 
family of phase-spaces, one for each topological type of
the Riemann surfaces. One of the crucial virtues of the 
Penner/Fock-coordinates is the fact that the Weil-Petersson symplectic
form has a particularly simple expression in these coordinates. 
The corresponding Poisson-brackets are in fact {\it constant} for
Fock's variables $z_e$,
\begin{equation}\label{poisson}
\{z_{e}^{},z_{e'}^{}\}=n_{e,e'}, \quad{\rm where}\quad
n_{e,e'}\in\{-2,-1,0,1,2\}.
\end{equation}
The value of $n_{e,e'}$ depends on how edges $e$ and $e'$ are imbedded
into a given fat graph. If $e$ and $e'$ don't have a common 
vertex at their ends, or if one of $e$, $e'$ starts and
ends at the same vertex then $n_{e,e'}=0$. In the case that 
$e$ and $e'$ meet at two vertices one has 
$n_{e,e'}=2$ (resp. $n_{e,e'}=-2$) 
if $e'$ is the first edge to the right\footnote{w.r.t.
to the orientation induced by the imbedding of the fat-graph 
into the surface} (resp. left) of $e$ at both
vertices, and $n_{e,e'}=0$ otherwise. 
In all the remaining cases $n_{e,e'}=1$ (resp. $n_{e,e'}=-1$) 
if $e'$ is the first edge to the right (resp. left) of $e$ at the common
vertex. 

If one considers a surface $\Sigma_g^s$ with genus $g$ and
$s$ boundary components
one will find $s$ central elements in the Poisson-algebra defined
by (\ref{poisson}). These central elements $c_k$, $k=1,\ldots, s$ 
are constructed as $c_k=\sum_{e\in E_k} z_e$,
where $E_k$ is the set of edges in the triangulation that emanates from the 
$k^{\rm th}$ boundary component. The value of $c_k$ 
gives the geodesic length of the $k^{\rm th}$ boundary component \cite{Fo}.

\subsection{Changing the triangulation}

Changing the triangulation amounts to a change of coordinates for 
the (decorated) Teichm\"uller spaces. Any two triangulations can be related
to each other by a sequence of the following elementary moves  called flips:\\
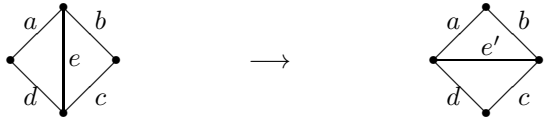
\begin{figure}[h]
\centering
\begin{picture}(200,40)
\put(20,0){\line(-1,1){20}}\put(5,3){$d$}
\put(20,0){\line(1,1){20}}\put(32,3){$c$}
\put(0,20){\line(1,1){20}}\put(5,32){$a$}
\put(40,20){\line(-1,1){20}}\put(32,32){$b$}
\put(20,0){\line(0,1){40}}\put(22,18){$e$}
\put(180,0){\line(-1,1){20}}\put(165,3){$d$}
\put(180,0){\line(1,1){20}}\put(192,3){$c$}
\put(160,20){\line(1,1){20}}\put(165,32){$a$}
\put(200,20){\line(-1,1){20}}\put(192,32){$b$}
\put(160,20){\line(1,0){40}}\put(178,22){$e'$}
\put(20,0){\circle*{3}}
\put(0,20){\circle*{3}}
\put(20,40){\circle*{3}}
\put(40,20){\circle*{3}}
\put(180,0){\circle*{3}}
\put(160,20){\circle*{3}}
\put(180,40){\circle*{3}}
\put(200,20){\circle*{3}}
\put(90,17){$\longrightarrow$}
\end{picture}
\caption{The elementary move between two triangulations}\label{fliplab}
\end{figure}

\begin{figure}[htb]
  \centering\setlength{\unitlength}{0.01in}
\begin{picture}(290,160)
\put(0,90){\begin{picture}(70,70)
\put(14,0){\line(-1,3){14}}
\put(56,0){\line(1,3){14}}
\put(0,42){\line(5,4){35}}
\put(70,42){\line(-5,4){35}}
\put(14,0){\line(1,0){42}}
\qbezier(14,0)(20,20)(35,70)
\qbezier(56,0)(50,20)(35,70)
\put(14,0){\circle*{3}}
\put(56,0){\circle*{3}}
\put(0,42){\circle*{3}}
\put(70,42){\circle*{3}}
\put(35,70){\circle*{3}}
\footnotesize
\put(14,35){$i$}
\put(33,17){$j$}
\put(55,35){$k$}
\end{picture}}
\put(110,90){\begin{picture}(70,70)
\put(14,0){\line(-1,3){14}}
\put(56,0){\line(1,3){14}}
\put(0,42){\line(5,4){35}}
\put(70,42){\line(-5,4){35}}
\put(14,0){\line(1,0){42}}
\put(56,0){\line(-4,3){56}}
\qbezier(56,0)(50,20)(35,70)
\put(14,0){\circle*{3}}
\put(56,0){\circle*{3}}
\put(0,42){\circle*{3}}
\put(70,42){\circle*{3}}
\put(35,70){\circle*{3}}
\footnotesize
\put(28,35){$i$}
\put(20,9){$j$}
\put(55,35){$k$}
\end{picture}}
\put(220,90){\begin{picture}(70,70)
\put(14,0){\line(-1,3){14}}
\put(56,0){\line(1,3){14}}
\put(0,42){\line(5,4){35}}
\put(70,42){\line(-5,4){35}}
\put(14,0){\line(1,0){42}}
\put(56,0){\line(-4,3){56}}
\put(0,42){\line(1,0){70}}
\put(14,0){\circle*{3}}
\put(56,0){\circle*{3}}
\put(0,42){\circle*{3}}
\put(70,42){\circle*{3}}
\put(35,70){\circle*{3}}
\footnotesize
\put(33,50){$i$}
\put(20,9){$j$}
\put(45,24 ){$k$}
\end{picture}}
\put(55,0){\begin{picture}(70,70)
\put(14,0){\line(-1,3){14}}
\put(56,0){\line(1,3){14}}
\put(0,42){\line(5,4){35}}
\put(70,42){\line(-5,4){35}}
\put(14,0){\line(1,0){42}}
\qbezier(14,0)(20,20)(35,70)
\put(14,0){\line(4,3){56}}
\put(14,0){\circle*{3}}
\put(56,0){\circle*{3}}
\put(0,42){\circle*{3}}
\put(70,42){\circle*{3}}
\put(35,70){\circle*{3}}
\footnotesize
\put(14,35){$i$}
\put(39,35){$j$}
\put(45,8){$k$}
\end{picture}}
\put(165,0){\begin{picture}(70,70)
\put(14,0){\line(-1,3){14}}
\put(56,0){\line(1,3){14}}
\put(0,42){\line(5,4){35}}
\put(70,42){\line(-5,4){35}}
\put(14,0){\line(1,0){42}}
\put(0,42){\line(1,0){70}}
\put(14,0){\line(4,3){56}}
\put(14,0){\circle*{3}}
\put(56,0){\circle*{3}}
\put(0,42){\circle*{3}}
\put(70,42){\circle*{3}}
\put(35,70){\circle*{3}}
\footnotesize
\put(33,50){$i$}
\put(26,24){$j$}
\put(45,8){$k$}
\end{picture}}
\put(35,70){$\searrow$}\put(43,74){\tiny$\omega_{jk}$}
\put(245,70){$\swarrow$}\put(235,74){\tiny$\omega_{jk}$}
\put(135,28){$\stackrel{\omega_{ij}}{\longrightarrow}$}
\put(80,118){$\stackrel{\omega_{ij}}{\longrightarrow}$}
\put(190,118){$\stackrel{\omega_{ik}}{\longrightarrow}$}
\end{picture}
\caption{The pentagon relation.}
\label{pentagon}
\end{figure}
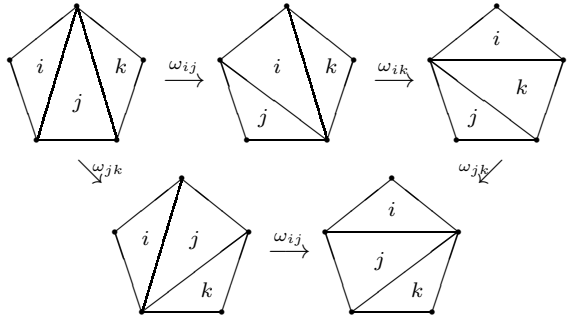

\noindent The change of variables corresponding to the elementary move
of Figure \ref{fliplab} is easy to describe: 
\begin{equation}\label{flipfovarclass}\begin{aligned}
z_a'=& z_a-\phi(-z_e),\\
z_d'=& z_d+\phi(+z_e),
\end{aligned} 
\quad z_e'=-z_e,\quad\begin{aligned}
z_b'=& z_b+\phi(+z_e),\\
z_c'=& z_c-\phi(-z_e),
\end{aligned}\quad{\rm where}\;\;\phi(x)=\ln(e^x+1),
\end{equation}
and all other variables are left unchanged.
These transformations generate a groupoid, the Ptolemy groupoid
\cite{Pe}, that
may be abstractly characterized by generators and relations:
One has a generator $\omega_{ij}$ 
whenever the triangles labeled
by $i$ and $j$ have an edge in common. The relation that 
characterizes the Ptolemy groupoid is called the Pentagon relation.
It is graphically represented in Figure \ref{pentagon}.

\subsection{The representation of the mapping class group}

\begin{figure}[h]
\epsfxsize7cm
\centerline{\epsfbox{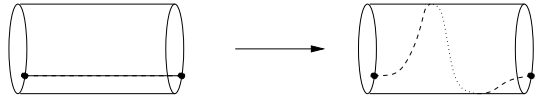}}
\caption{Action of a Dehn-twist on an annulus.}\label{dehn}
\end{figure}

The mapping class group ${\rm MC}(\Sigma)$ consists of
diffeomorphisms of the Riemann surface $\Sigma$ which are
not isotopic to the identity. It is generated by the Dehn-twists, 
which act on an annulus $A$ in the way indicated in Figure 
\ref{dehn}. 

Elements ${\rm MC}(\Sigma)$ will map any
graph drawn on the surface $\Sigma$, in particular
any triangulation of $\Sigma$, into another one. Since any
two triangulations can be connected by a sequence of elementary moves
one may represent any element ${\rm MC}(\Sigma)$ by the 
corresponding sequence of flips. This means that the 
mapping class group ${\rm MC}(\Sigma)$ is a subgroup of 
the Ptolemy groupoid ${\rm Pt}(\Sigma)$.

It is extremely useful to think of
the algebraically complicated mapping class group ${\rm MC}(\Sigma)$ 
as being embedded into the Ptolemy groupoid.

\section{Quantization of Teichm\"uller spaces}

\subsection{Algebra of observables and Hilbert space}
The simplicity of the Poisson brackets (\ref{poisson}) makes part of the
quantization quite simple. To each edge $e$ of a triangulation 
of a Riemann surface $\Sigma_g^s$ associate a quantum operator 
$\sz_e$. The algebra of observables $\CA(\Sigma_g^s)$ 
will be the algebra with generators $\sz_e$, relations 
\begin{equation}\label{comm}
[\sz_{e}^{},\sz_{e'}^{}]=2\pi i b^2 \{z_{e}^{},z_{e'}^{}\}, 
\end{equation}
and hermiticity assignment $\sz_e^{\dagger}=\sz_e^{}$. 
The algebra $\CA(\Sigma_g^s)$ 
has a center with generators
$\sfc_k$, $k=1,\ldots, s$ defined by $\sfc_k=\sum_{e\in E_k}\sz_e$,
where $E_k$ is the set of edges in the triangulation that emanates from the 
$k^{\rm th}$ boundary component. 
The representations of $\CA(\Sigma_g^s)$ that we are going to consider
will therefore be such that the generators $\sfc_k$ are 
represented as the operators of multiplication by real 
positive numbers $l_k$. Geometrically one may interpret 
$l_k$ as the geodesic length of the 
$k^{\rm th}$ boundary component \cite{Fo}. 
The tuple $\Lambda=(l_1,\dots,l_s)$ of lengths of the boundary components
will figure as a label of the representation 
$\pi(\Sigma_g^s,\Lambda)$ of the algebra $\CA(\Sigma_g^s)$.
   
To complete the definition of the 
representation $\pi(\Sigma_g^s,\Lambda)$ by operators 
on a Hilbert space $\CH(\Sigma_g^s)$ one just needs to find 
linear combinations $\sq_1,\dots,\sq_{3g-3+s}$ and 
$\spp_1,\dots,\spp_{3g-3+s}$ of the 
$\sz_{e}$ that satisfy $[\spp_m,\sq_n]=(2\pi i)^{-1}\delta_{mn}$.
The representation of $\CA(\Sigma_g^s,\Lambda)$ on 
$\CH(\Sigma_g^s):=L^2(R^{3g-3+s})$ is  
defined by choosing 
the usual Schr\"odinger representation for the $\sq_i$, $\spp_i$.  

Let us discuss the example of a sphere with four holes. 
We shall consider the fat graph drawn in Figure \ref{fourpt} below.
\begin{figure}[h]
\epsfxsize7cm
\centerline{\epsfbox{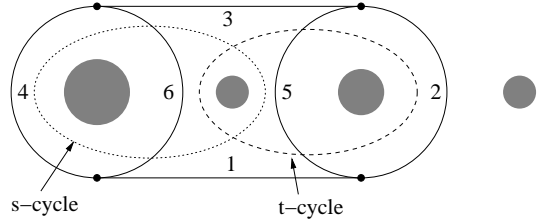}}
\caption{Fat graph for the sphere with four holes (shaded) 
with numbered edges.}
\label{fourpt}\end{figure}
The algebra $\CA(\Sigma_{0}^{4})$
has six generators $\sz_i$ $i=1,\dots 6$ with nontrivial relations
$[\sz_i,\sz_j]=2\pi i b^2$ for 
\begin{equation}
\begin{aligned}
(i,j)\;\in\;\{ & 
(1,2)\; ,\; (1,6) \; ,\; (2,3)\; ,\;(3,4)\; ,\;
 (3,5)\; , \; (4,1)\; ,\; (5,1)\; ,\;(6,3)\;\}.
\end{aligned}
\end{equation}
The four central elements corresponding to the holes in $\Sigma_{0,4}$ are 
\begin{equation}
\begin{aligned}
\sfc_1\;=\;& \sz_4+\sz_6,\\
\sfc_3\;=\;& \sz_2+\sz_5,
\end{aligned}\qquad\begin{aligned}
\sfc_2\;=\;& \sz_1+\sz_3+\sz_5+\sz_6,\\
-\sfc_4\;=\;& \sz_1+\sz_2+\sz_3+\sz_4.
\end{aligned}
\end{equation}
After fixing the lengths of the four holes one is left with two variables,
say $\sz_1$ and $\sz_5$. Choosing the Schr\"odinger
representation for $\sz_1$, $\sz_5$ one simply finds $\CH(\Sigma_0^4)
\simeq L^2(\BR)$.

\subsection{Representation of the mapping class group on $\CA(\Sigma)$ 
}

The first task is to find the quantum counterparts of the 
changes of variables corresponding to the elements
of the Ptolemy groupoid. A handy formulation of 
the solution \cite{Fo,CF} can be given in terms 
of the Fock-variables: The change of variables
corresponding to the
elementary move depicted in Figure \ref{fliplab} is given by
\begin{equation}\label{flipfovar}\begin{aligned}
\sz_a'=& \sz_a-\phi_{b}(-\sz_e),\\
\sz_d'=& \sz_d+\phi_{b}(+\sz_e),
\end{aligned} 
\qquad \sz_e'=-\sz_e,\qquad\begin{aligned}
\sz_b'=& \sz_b+\phi_{b}(+\sz_e),\\
\sz_c'=& \sz_c-\phi_{b}(-\sz_e),
\end{aligned}
\end{equation}
where the special function $\phi_{b}(x)$ 
is defined as
\begin{equation}
\phi_b(z)\;=\;\frac{\pi b^2}{2}
\int\limits_{i0-\infty}^{i0+\infty}dw
\frac{e^{-i zw}}{\sinh(\pi w)
\sinh(\pi b^{2}w)}.
\end{equation}
$\phi_{b}(x)$ represents the quantum deformation of the
classical expression given in (\ref{flipfovarclass}).
The formulae (\ref{flipfovar}) define a representation of the Ptolemy groupoid
by automorphisms of $\CA(\Sigma)$ \cite{CF}.

Let us now recall that the mapping class group ${\rm MC}(\Sigma)$ 
can be embedded
into the Ptolemy groupoid. Having realized the latter therefore 
gives a representation of ${\rm MC}(\Sigma)$ by automorphisms of 
$\CA(\Sigma)$.

\subsection{Representation of the mapping class group on $\CH(\Sigma)$}

The next problem is to define unitary operators that generate 
the action of the mapping class group on $\CH(\Sigma)$ \cite{Ka1,Ka2}. 

\begin{figure}[h]
\epsfxsize6cm
\centerline{\epsfbox{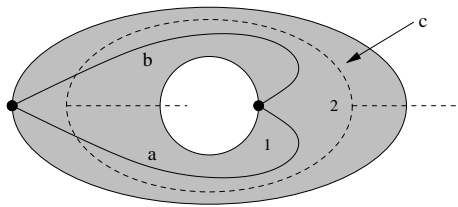}}
\caption{Triangulation of an annulus and the dual fat-graph (dashed).}
\label{trann}\end{figure}

In the case that $g\geq 2$ is is possible to describe the 
action of the generators rather simply. For this case it is known
\cite{Ge} that the mapping class group is generated by 
the Dehn-twists along {\it nonseparating}\footnote{Cutting along
such curves preserves connectedness of $\Sigma$, it just 
opens a handle.}  
simple closed curves. In an annular neighborhood of such a
curve $c$ one may always bring the triangulation into the form depicted
in Figure \ref{trann}.
In this case it suffices to consider the two variables $\sz_a$, $\sz_b$
associated to the edges $a$ and $b$ marked in Figure \ref{trann}.
They satisfy the algebra $[\sz_b,\sz_a]=4\pi i b^2$. The Dehn twist
can be represented by a simple flip. It can be checked that
the Dehn twist around $c$ on $\CH(\Sigma)$ is represented
by the following operator $\SD_c$:
\begin{equation}\label{dehngen}
\SD_c\;=\;e^{2\pi i(\spp^2-c_b^2)}\,{\rm e}_b(\spp-\sx),
\qquad\begin{aligned} 2\pi b \sx \;=&\;\fr{1}{2}(\sz_b-\sz_a), \\
2\pi b \spp\;=&\;\fr{1}{2}(\sz_a+\sz_b).
\end{aligned}
\end{equation}
where the special function $e_b(x)$ can be  
defined in the strip 
$|\Im z|<|\Im c_b|$, $c_b\equiv i(b+b^{-1})/2$ by means of the 
integral representation
\begin{equation}
\log e_b(z)\;\equiv\;\frac{1}{4}
\int\limits_{i0-\infty}^{i0+\infty}\frac{dw}{w}
\frac{e^{-i 2 zw}}{\sinh(bw)
\sinh(b^{-1}w)}.
\end{equation}
Formulae (\ref{dehngen}) and (\ref{flipfovar})  
suffice to completely define the
action of ${\rm MC}(\Sigma_g^s)$ on $\CH(\Sigma_g^s)$ for $g>1$: 
By means of 
(\ref{flipfovar})  
one may construct the representation of the Dehn twist along 
$g$ in an arbitrary triangulation from (\ref{dehngen}).

The operators that represent ${\rm MC}(\Sigma_g^s)$
are of course defined by their action on $\CA(\Sigma_g^s)$
only up to multiplication by phase factors. Controlling the 
possible occurrence of phase factors in the relations of the 
mapping class group is a subtle task. 
An elegant formalism for handling this problem
was given by Kashaev in \cite{Ka1}. It
uses an enlarged set of variables, where
pairs of variables are associated to the 
$2M=4g-4+2s$ triangles of a triangulation instead
of its edges. The reduction to $(\CA(\Sigma_g^s),\CH(\Sigma_g^s))$ 
can be described
with the help of a simple set of constraints \cite{Ka1}. 
It indeed turns out that
the mapping class group is realized only projectively \cite{Ka2}.

\section{Liouville theory in the Moore-Seiberg formalism}

We shall now briefly review the construction of the conformal blocks 
in quantum Liouville theory \cite{TL,TL2} from a geometric perspective.
The result will be another assignment 
$\Sigma\rightarrow (\CH^{\rm L}(\Sigma),\pi^{\rm L}(\Sigma))$, 
where $\CH^{\rm L}(\Sigma)$ is the space of conformal blocks in Liouville
theory and $\pi^{\rm L}(\Sigma)$ is the representation of the mapping class
group on $\CH^{\rm L}(\Sigma)$.

\subsection{Holomorphic factorization}

The spectrum of quantum Liouville theory can be represented as follows:
\begin{equation}\label{spec}
\CH^{\rm L}\;\simeq\; \int_{\BS}d\alpha \;\CV_{\alpha}\otimes\bar{\CV}_{\alpha}
,\qquad\BS\equiv \frac{Q}{2}+i\BR^+.
\end{equation}
In (\ref{spec})
we used the notation $\CV_{\alpha}$, $\bar{\CV}_{\alpha}$ for the  
unitary highest weight representations   
of the Virasoro algebras 
which are generated from the modes of the holomorphic and 
anti-holomorphic parts
of the energy-momentum tensor respectively. 
The central charge of the representations is given by the parameter
$b$ via $c=1+6Q^2$, $Q=b+b^{-1}$, and the
highest weight of the representations $\CV_{\alpha}$, $\bar{\CV}_{\alpha}$ is 
parameterized as $\Delta_{\alpha}=\alpha(Q-\alpha)$.

Quantum Liouville theory in genus zero is fully characterized by the 
set of n-point functions of the primary fields $V_{\alpha}(z,\bar{z})$,
which are the quantized exponential functions $e^{2\alpha\phi(z,\bar{z})}$
of the Liouville field. The $V_{\alpha}(z,\bar{z})$ are primary fields
with conformal weights $(\De_{\al},\De_{\al})$. As usual 
\cite{BPZ} one may consider $V_{\alpha}(z,\bar{z})$ as the generator
of a family of
descendant fields $V_{\alpha}(v\ot w\,|\,z,\bar{z})$
parameterized by vectors $v\ot w\in
\CV_{\alpha}\otimes\bar{\CV}_{\alpha}$.

In principle it should always be possible
to evaluate an n-point function
\[
\langle V_{\alpha_s}(z_s,\bar{z}_s)\dots V_{\alpha_1}(z_1,\bar{z}_1)
\rangle
\]  
by inserting complete sets of intermediate states between each 
pair of fields $V_{\alpha_i}(z_i,\bar{z}_i)$ and 
$V_{\alpha_{i+1}}(z_{i+1},\bar{z}_{i+1})$. 
Due to the factorized structure of the Hilbert space
(\ref{spec}) one finds a holomorphically factorized form for the 
n-point functions, which may be written as
\newcommand{\sst}{\scriptscriptstyle}
\newcommand{\ti}{\times}
\[ \begin{aligned}
\big\langle \,V_{\alpha_s}(v_s\ot w_s\,| & \,z_s,\bar{z}_s) 
\dots V_{\alpha_1}( v_1\ot w_1\,|\,z_1,\bar{z}_1)
\, \big\rangle\;=\;\\
& \;=\; \int\limits_{\BS_{s}} dS \;C(S|E)\,
\CF_{\rm\sst S,E}^{\Sigma}\big( v_s\ot\dots\ot v_1\big)
\bar{\CF}_{\rm\sst S,E}^{\Sigma}\big( w_s\ot\dots\ot w_1\big),
\end{aligned} \] 
where we have used the following notation. 
The tuple ${\rm E}=(\alpha_s,\dots,\alpha_1)$ represents the ``external''
parameters, whereas 
\[
S=(\be_{s-3},\dots,\be_1)\in\BS_s\equiv\BS^{s-3}
\]
comprises the variables of integration. 
The fact that $\CF_{\rm\sst S,E}^{\Sigma}$
(resp. $\bar{\CF}_{\rm\sst S,E}^{\Sigma}$) depends holomorphically
(resp. antiholomorphically) on $ z_1,\dots, z_s$ is indicated via
$\Sigma\equiv \CP^1\setminus\{ z_1,\dots, z_s\}$.
The conformal blocks 
$\CF_{\rm\sst S,E}^{\Sigma}\big( v_s\ot\dots\ot v_1\big)$ 
are key objects. 

\subsection{The conformal Ward identities}

It is well-known that the conformal blocks are strongly 
constrained by the conformal Ward-identities which express
the conservation of energy-momentum on the Riemann-sphere
${\mathbb P}^1\setminus\{z_1,\dots,z_s\}$. 

Mathematically speaking a conformal block is a functional
\[
\CF_{\rm\sst E}^{\Sigma} :
\CV_{\al_s}\ot\dots\ot\CV_{\al_1}\rightarrow \BC\] 
that satisfies the 
following invariance condition. Let $v(z)$ be a 
meromorphic vector field that is
holomorphic on ${\mathbb P}^1\setminus\{z_1,\dots,z_s\}$.
Write the Laurent-expansion of $v(z)$ 
in an annular neighborhood of $z_k$ 
in the form $v(z)=\sum_{n\in\BZ}v_n^{(k)}(z-z_k)^{n+1}$, and define
an operator $T[v]$ on $\CV_{\al_s}\ot\dots\ot\CV_{\al_1}$
by 
\newcommand{\id}{{\rm id}}
\[
T[v]\;=\;\sum_{k=1}^s\sum_{n\in\BZ}v_n^{(k)}L_n^{(k)},\qquad
L_n^{(k)}=\id\ot\dots\ot
\underset{\scriptstyle\rm (k-th)}{L_n}\ot\dots\ot\id.
\] 
The conformal Ward identities can then be formulated as 
the condition that 
\begin{equation}\label{Wardid}
\CF_{\rm\sst E}^{\Sigma}\big(\,T[v] w\,\big)\;=\;0 
\end{equation}
holds for all $ w\in  \CV_{\al_s}\ot\dots\ot\CV_{\al_1}$ and all
meromorphic vector fields $v$ that are holomorphic 
on ${\mathbb P}^1\setminus\{z_1,\dots,z_s\}$. 

By choosing vector fields $v$ 
that are singular at a single point only one 
recovers the usual rules for moving Virasoro generators from one puncture 
to another. In this way one may convince oneself that the conformal
block is uniquely determined by the value 
$\CF_{\rm\sst E}^{\Sigma}(v_{\rm\sst E})\in\BC$ that it takes 
on the product of highest weight states
$v_{\rm\sst E}\equiv v_{\al_s}\ot\dots\ot v_{\al_1}$.

A concrete representation for the  
conformal blocks in genus zero
can be given by means of the chiral
vertex operators $\sh_{\al_3\al_1}^{\al_2}(z)$ constructed in refs.
\cite{TL,TL2}, and their descendants $\sh_{\al_3\al_1}^{\al_2}(v\,|\,z)$,
$v\in\CV_{\al}$. Let $\Sigma\equiv \CP^1\setminus\{ z_1,\dots, z_s\}$,
${\rm E}=(\alpha_s,\dots,\alpha_1)$, ${\rm S}=(\be_{s-3},\ldots,\be_1)$.
\begin{equation}\begin{aligned}
 \CH_{\rm\sst S,E}^{\Sigma} & \big( v_s\ot\dots\ot v_1\big)\;=\;\\
& \big\langle\,v_0\, , \,
\sh_{0,\al_s}^{\al_s}( v_s\,|\,z_s)\sh_{\al_s\be_{s-3}}^{\al_{s-1}}
( v_{s-1}\,|\,z_{s-1})\dots
\sh_{\be_1\al_1}^{\al_{2}}( v_2\,|\,z_{2})\sh_{\al_1,0}^{\al_1}
( v_1\, |\,z_1) v_0
\big\rangle
\end{aligned}
\end{equation}
where $v_0$ is the highest weight vector in the vacuum representation
$\CV_0$.

It is well-known that the space of solutions to the 
condition (\ref{Wardid}) is one-dimensional for 
the case of the three-punctured sphere $s=3$. Invariance under global
conformal transformations allows one to assume that
$\Sigma_3=\BP^1\setminus\{0,1,\infty\}$. 
We will adopt the normalization from \cite{TL} and denote
$\CC^{\Sigma_3}_{\rm\sst E}$ the unique conformal block 
that satisfies 
$\CC^{\Sigma_3}_{\rm\sst E}(v_{\al_3}\ot v_{\al_2}\ot v_{\al_1})=
{\rm N}(\al_3,\al_2,\al_1)$. The function ${\rm N}(\al_3,\al_2,\al_1)$
is defined in \cite{TL} but will not be needed in the following.

Let us furthermore note that the case of $s=2$
corresponds to the invariant bilinear form 
\newcommand{\bra}{\langle}
\newcommand{\ket}{\rangle}
$\bra .\, , .\ket_{\al}:
\CV_{\al}\ot\CV_{\al}\rightarrow \BC$ which is defined 
such that $\bra L_{-n} w, v\ket_{\al}=
\bra  w,L_{n} v\ket_{\al}$. 

\subsection{Sewing of conformal blocks}

For $n>3$ one
may generate large classes of solutions 
of the conformal Ward identities by the 
following ``sewing'' construction. Let 
$\Sigma_i$, $i=1,2$ be Riemann surfaces with $m_i+1$ punctures, 
and let $\CG_{\rm\sst E_2}^{\Sigma_2}$ and $\CH_{\rm\sst E_1}^{\Sigma_1}$
be conformal 
blocks associated to $\Sigma_i$, $i=1,2$ and representations
labeled by $E_2=(\al_{m_2},\dots,\al_1,\al)$ 
and $E_1=(\al,\al'_{m_1},\dots,\al'_{1})$ respectively.
Let $p_i$, $i=1,2$ be the distinguished punctures on $\Sigma_i$
that are associated to the representation $\CV_{\al}$. Around 
$p_i$ choose local coordinates $z_i$ such that $z_i=0$ parameterizes
the points $p_i$ themselves. Let $A_i$ be the annuli 
$r<|z_i|<R$, and $D_i$ be the disks $|z_i|\leq r$.
We assume $R$ to be small enough such
that the $A_i$ contain no other punctures.  
The surface $\Sigma_2\infty\Sigma_1$ 
that is obtained by ``sewing'' $\Sigma_2$ and $\Sigma_1$ will
be 
\[
\Sigma_2\,\infty\,\Sigma_1\;=\;\big(
(\Sigma_2\setminus D_2)\cup
(\Sigma_1\setminus D_1)\big)\,/\,{\sim},
\]
where $\sim$ denotes the identification of annuli $A_2$ and $A_1$
via $z_1z_2=rR$. The conformal block 
$\CF_{\al,{\rm \sst E_{21}}}^{\Sigma_2\infty\Sigma_1}$ assigned to
$\Sigma_2\infty\Sigma_1$ , 
${\rm E}_{21}=(\al_{m_2},\dots,\al_1,\al'_{m_1},\dots,\al'_{1})$
and $\al\in\BS$ will then be
\begin{equation}\begin{aligned}
{}& \CF_{\al,{\rm \sst E_{21}}}^{\Sigma_2\infty\Sigma_1} 
\big(v_{m_2}\ot\dots\ot v_{1}\ot w_{m_1}\ot\dots \ot w_{1}\big)
\;=\;\\
 &=\sum_{\imath,\jmath\in\BI}\;
\CG_{\rm \sst E_2}^{\Sigma_2} 
\big( v_{m_2}^{}\ot\dots\ot v_{1}^{}\ot v_{\imath}^{}\big)
\,\bra  v^{\vee}_{\imath},e^{-tL_0} v_{\jmath}^{}
\ket_{\al}^{\phantom{\dagger}}\,
\CH_{\rm \sst E_1}^{\Sigma_1} 
\big( v^{\vee}_{\jmath}\ot w_{m_1}^{}\ot\dots\ot w_{1}^{}\big).
\end{aligned}
\end{equation}
The sets $\{ v_{\al,\imath};\imath\in\BI\}$ and
$\{ v_{\al,\imath}^{\vee};\imath\in\BI\}$ 
are supposed to represent mutually
dual bases for $\CV_{\al}$ in the sense that $
\bra  v_{\al,\imath}^{\phantom{\vee}}, v_{\al,\jmath}^{\vee}
\ket_{\al}^{\phantom{\vee}}=\de_{\imath\jmath}$. In a similar way one 
may construct the conformal blocks associated to a surface that
was obtained by sewing two punctures on a single Riemann surface.

\subsection{Feynman rules for the construction of conformal blocks}

The sewing construction allows one to construct large classes of 
solutions to the conformal Ward identities from simple pieces. 
The resulting construction resembles
the construction of field theoretical amplitudes by 
the application of a set of Feyman rules.
Let us summarize the basic ingredients and their geometric 
counterparts. \\[1ex]
\newcommand{\gs}{$\frac{\quad}{}\;\;$}
{\sc Propagator} \gs Invariant bilinear form:
\[
\bra v_2\,,\,e^{-tL_0}v_1\ket_{\CV_{\al}}^{}\quad \sim\quad
\lower.8cm\hbox{\epsfig{figure=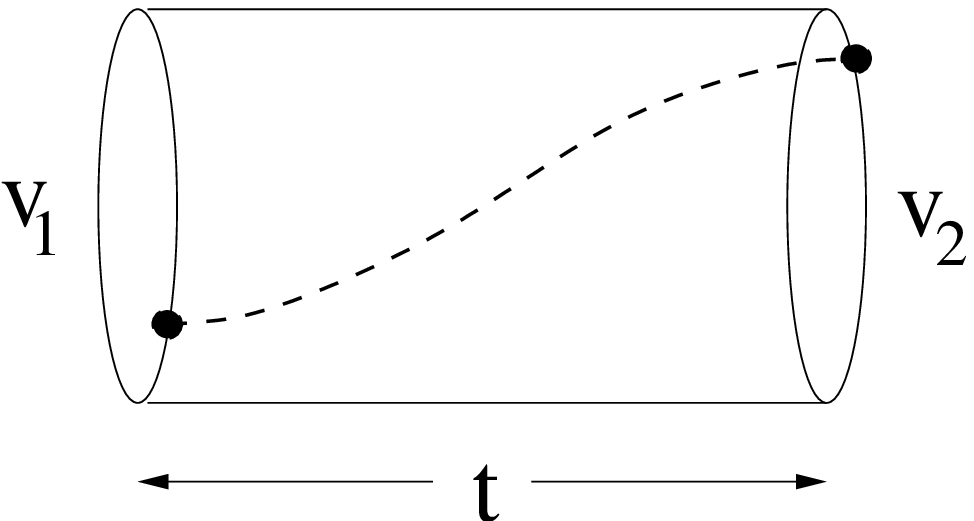,height=1.6cm}}
\]
{\sc Vertex} \gs Invariant trilinear form:
\[
\CC^{\Sigma_3}_{\rm\sst E}(v_3,v_2,v_1)\quad \sim\quad
\lower1.5cm\hbox{\epsfig{figure=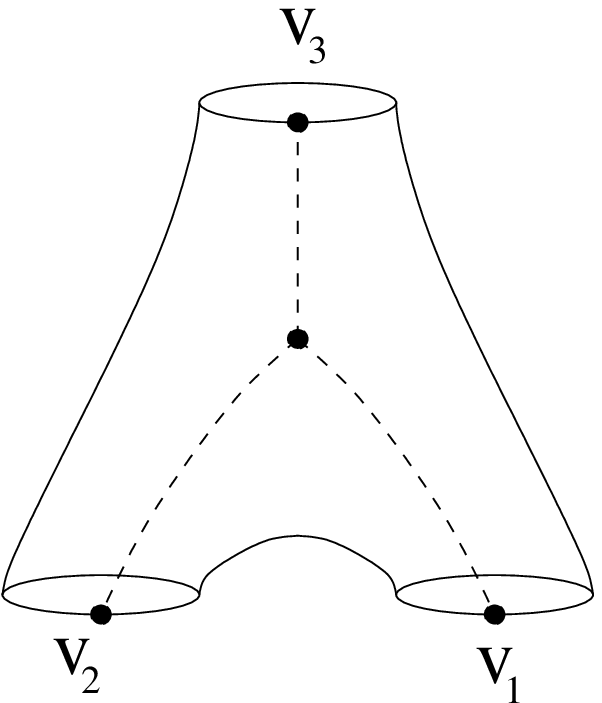,height=3cm}}
\]
{\sc Gluing} \gs Completeness: 
\[\begin{aligned}{}&
\lower0.7cm\hbox{\epsfig{figure=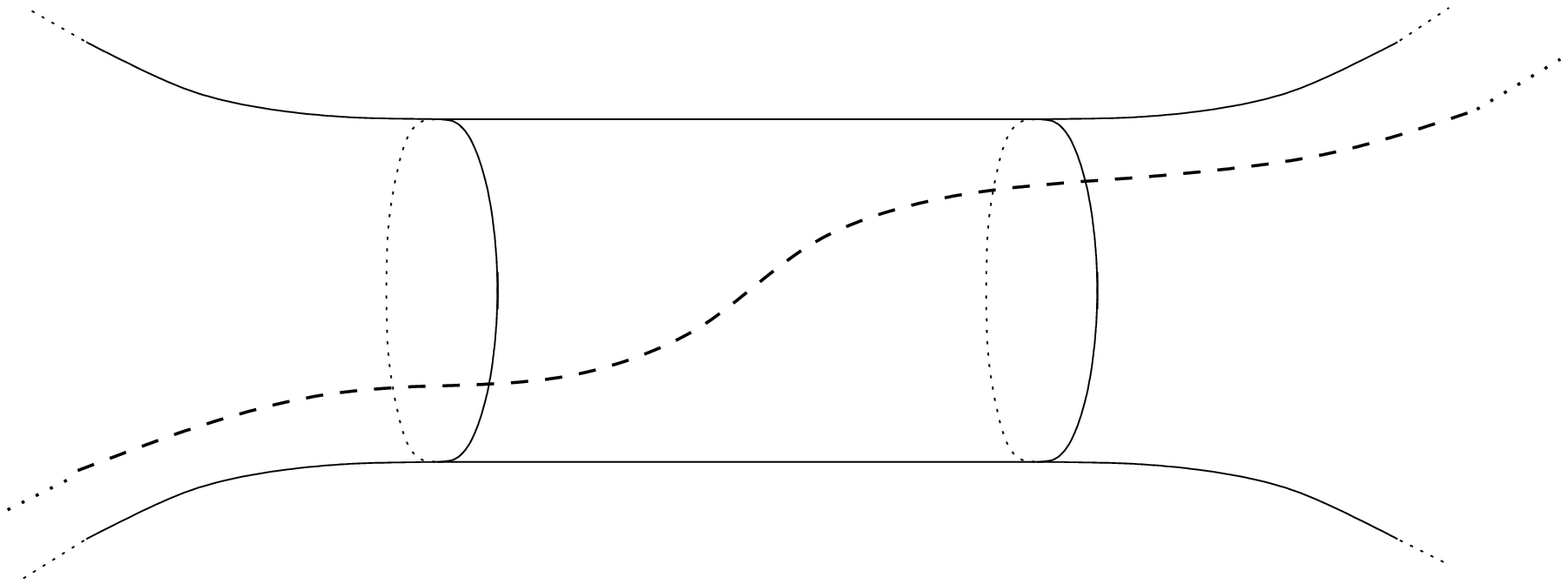,height=1.4cm}} \;=
\;\sum_{\imath,\jmath\in\BI}
\left(\,\;\lower0.7cm\hbox{\epsfig{figure=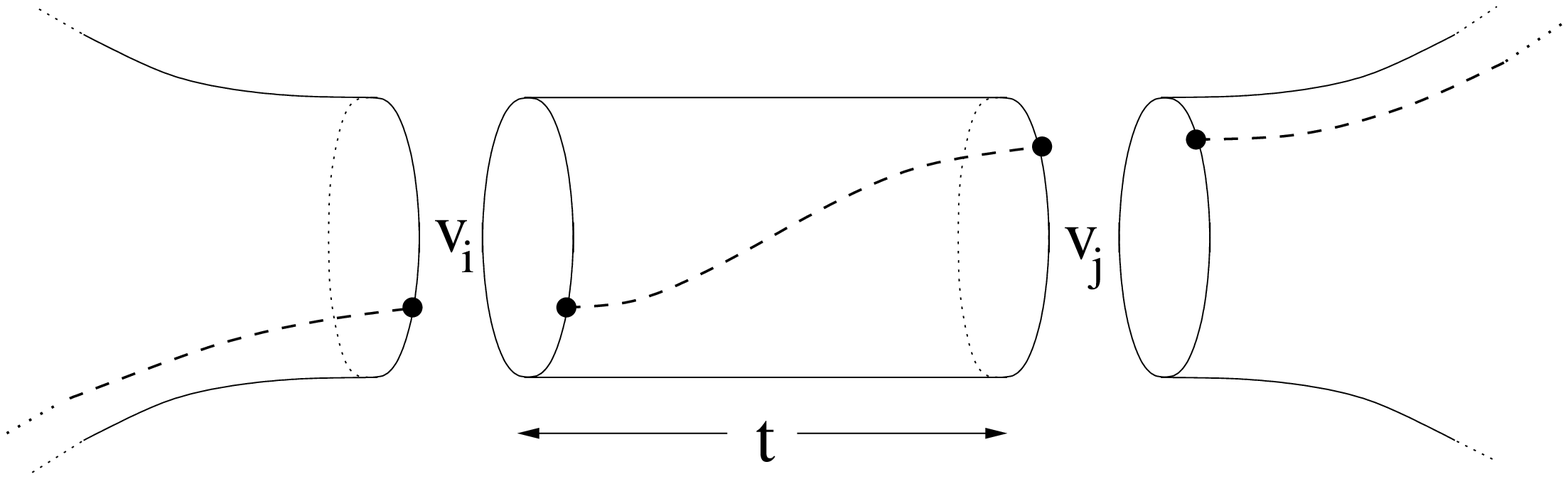,height=1.4cm}}\;\,\right)
\end{aligned}
\]
The dashed lines have been introduced to take care of the 
fact that the rotation of a boundary circle by $2\pi$ (Dehn twist)
is not represented 
trivially. It acts by multiplication with $e^{2\pi i \De_{\al}}$. 
This describes a part of the action of the mapping class group
on the spaces of conformal blocks. The Riemann surfaces that
are obtained by gluing cylinders and three-holed spheres as drawn will 
therefore carry a trivalent graph which we will call 
Moore-Seiberg graph. These graphs are not to be confused with the fat graphs
that we had encountered in the previous subsections. 
 
The gluing construction furnishes  spaces of conformal blocks 
$\CH^{\rm L}(\Sigma,\Gamma)$ 
associated to 
a Riemann surface $\Sigma$ together with
a Moore-Seiberg graph $\Gamma$. A basis for this space is obtained
by coloring the ``internal'' edges of the Moore-Seiberg graph $\Gamma$
with elements of $\BS$, for example
\[
\CH^{\rm L}\Biggl(
\lower.4cm\hbox{\epsfig{figure=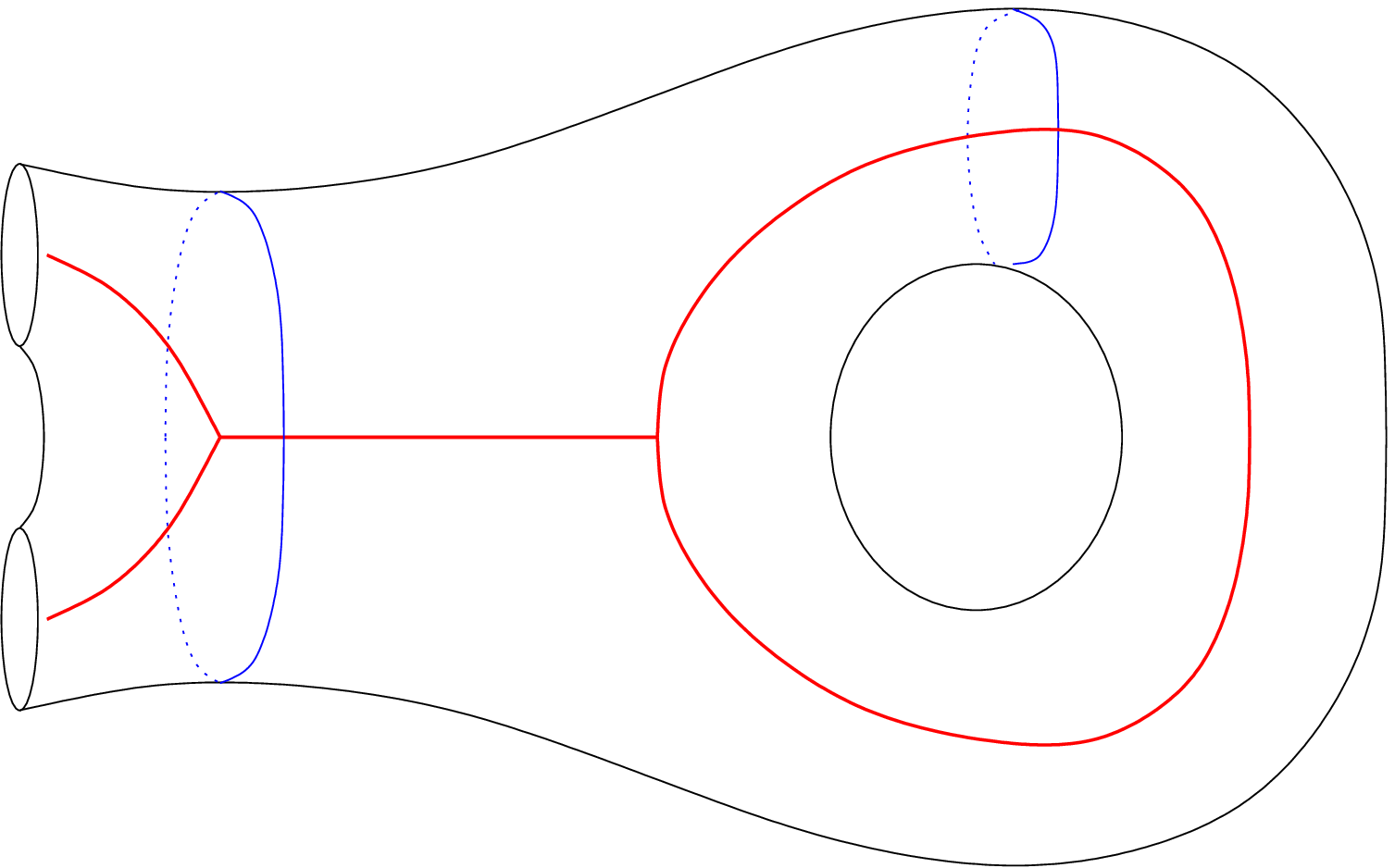,height=1cm}}
\Biggr)\;= \;
{\rm Span}\left\{\;\;
\lower0.9cm\hbox{\epsfig{figure=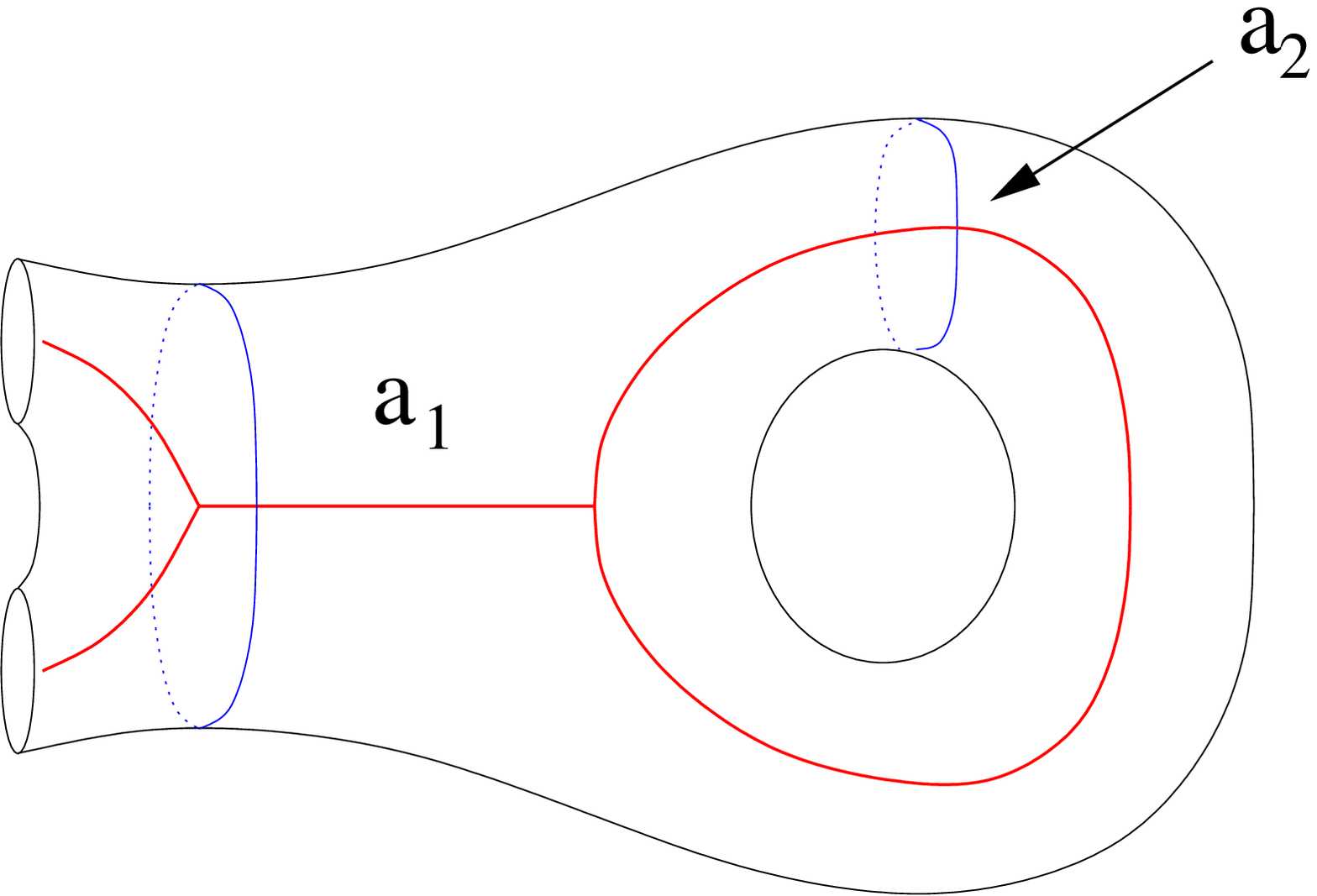,height=2.1cm}}
\;\; ;\; \al_2, \al_1\in\BS\;\right\}.
\]
\newcommand{\ra}{\rightarrow}
In order to show that the Hilbert spaces associated to 
each two Moore-Seiberg graphs $\Gamma_1$ and $\Gamma_2$ are 
isomorphic,
$\CH^{\rm L}(\Sigma,\Gamma_1)\simeq
\CH^{\rm L}(\Sigma,\Gamma_2)\simeq
\CH^{\rm L}(\Sigma),$
one needs to find unitary operators 
$
\SU_{\Gamma_2\Gamma_1}^{}:
\CH^{\rm L}(\Sigma,\Gamma_1)\;\ra\;\CH^{\rm L}(\Sigma,\Gamma_2).$

\subsection{The Moore-Seiberg groupoid}

The transitions from one Moore-Seiberg graph
to another generate another groupoid that will be called the
Moore-Seiberg groupoid. 
A set of generators is pictorially represented 
in Figures \ref{MS1}-\ref{MS3} below. 
\begin{figure}[h]
\centerline{\epsfxsize7cm\epsfbox{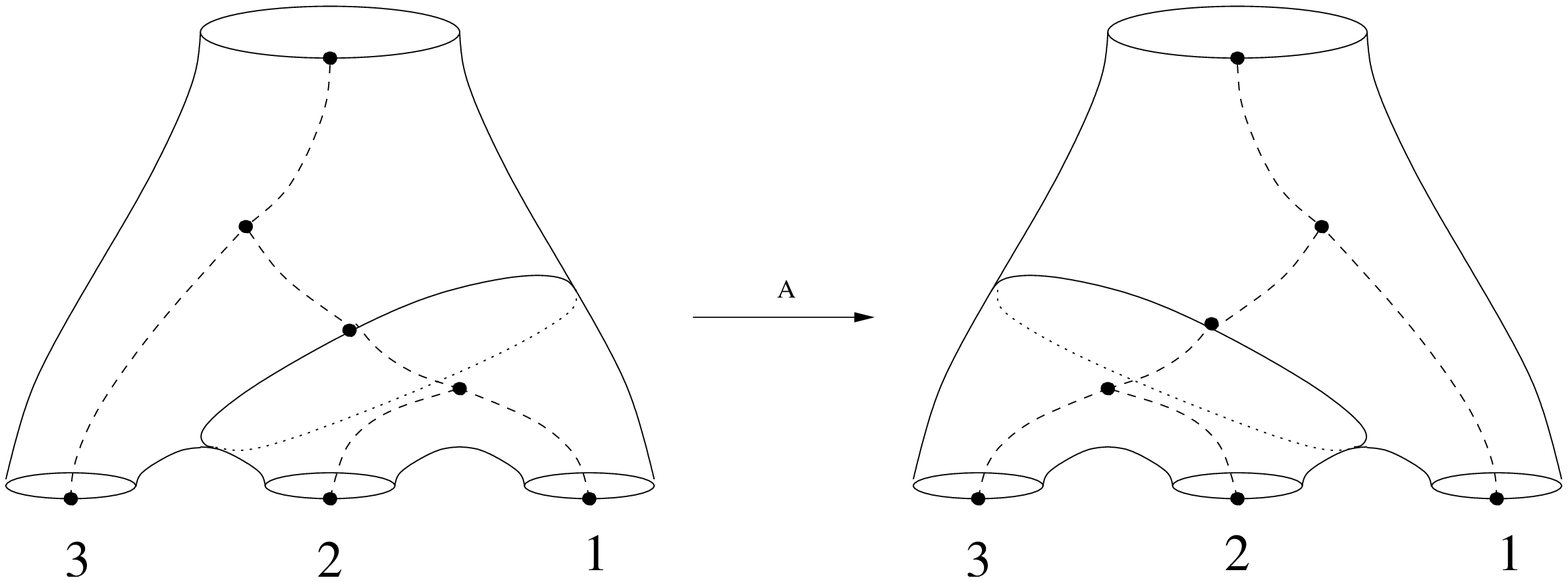}} 
\caption{A-move}\label{MS1}\end{figure}
\begin{figure}[h]
\centerline{\epsfxsize7cm\epsfbox{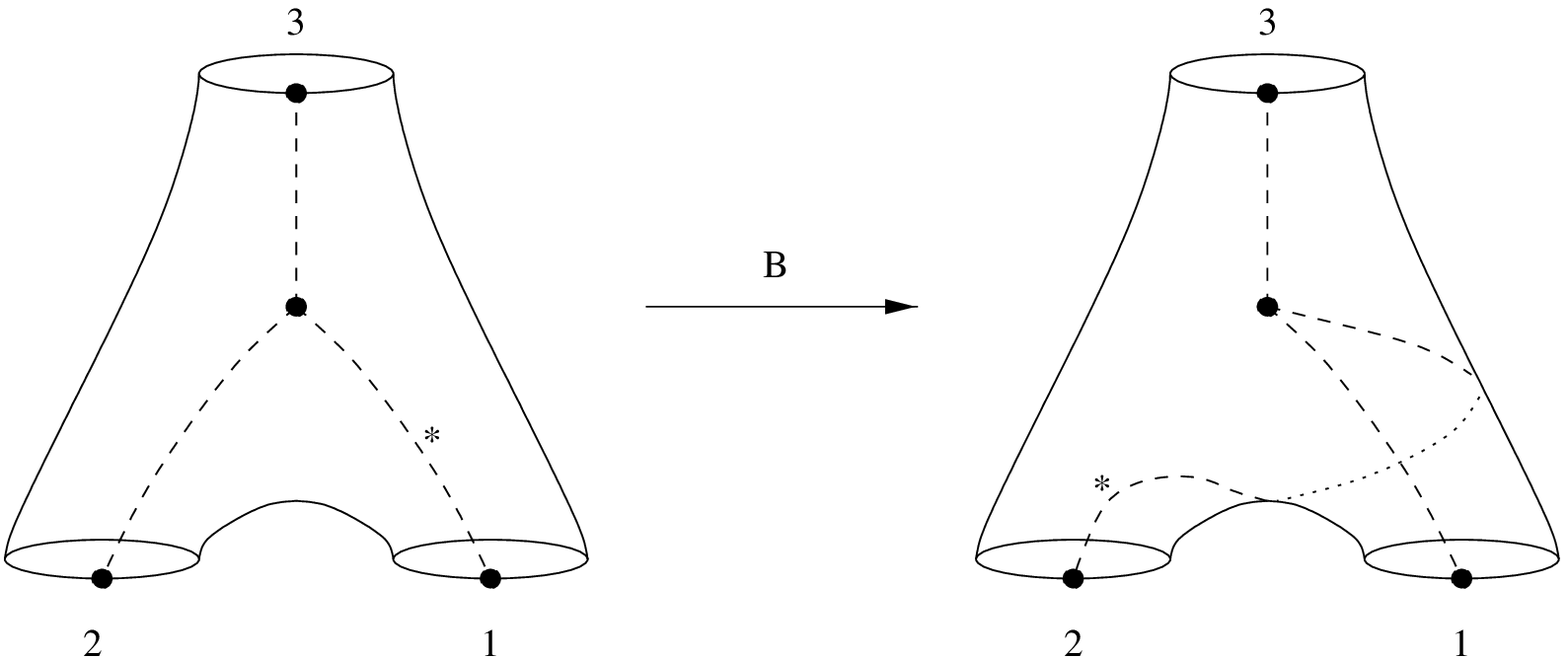}} 
\caption{B-move}\end{figure}
\begin{figure}[h]
\centerline{\epsfxsize7cm\epsfbox{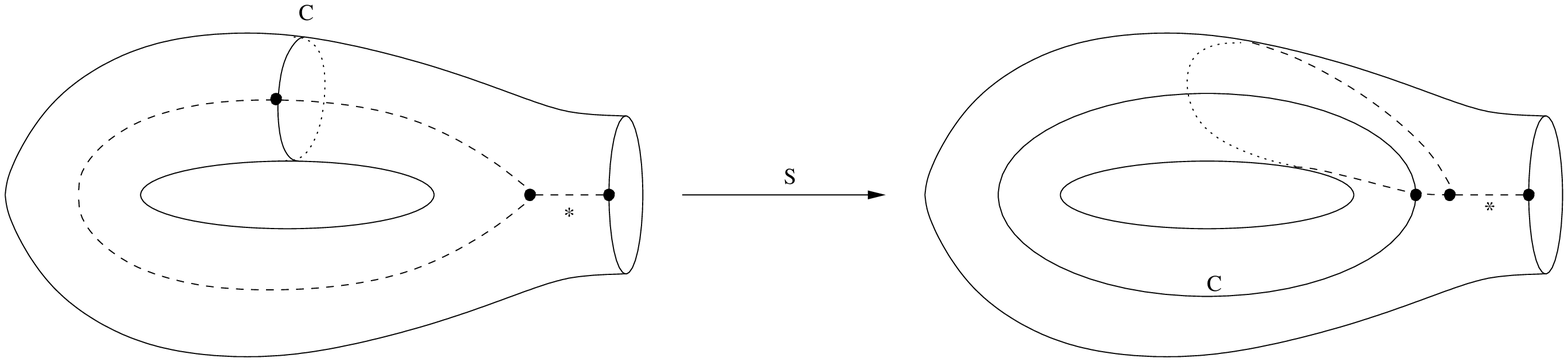}}
\caption{S-move}\label{MS3} \end{figure}

The
Moore-Seiberg groupoid can also be characterized in terms of the generators
depicted in Figures \ref{MS1}-\ref{MS3} and certain relations, but
the set of relations for the 
Moore-Seiberg-groupoid is more complicated than the one 
for the Ptolemy-groupoid \cite{MS1,BK}. 

\subsection{Representation of the Moore-Seiberg 
groupoid on $\CH^{\rm L}(\Sigma)$} 

In order to characterize a 
representation of the Moore-Seiberg groupoid it suffices 
to find the operators 
$\SU_{\Gamma_2\Gamma_1}^{}$ for the cases where $\Gamma_2$ and $\Gamma_1$
differ by an A-, B- or S-move. In the case of the Liouville 
conformal blocks in genus zero this was done in \cite{TL,TL2}.\\[1ex]
{\sc A-Move}: In order to describe the representation of the
A-move let $\Sigma$ be the 
four-punctured sphere, with parameters ${\rm E}=(\al_4,\dots,\al_1)$
associated to the four punctures respectively. The conformal blocks
corresponding to the sewing patterns indicated on the
left and right halfs of Figure \ref{MS1} will be denoted
$\CF_{\rm\sst E,\al_s}^{\Sigma}$ and $\CG_{\rm\sst E,\al_t}^{\Sigma}$
respectively. The A-move is then represented as an integral
transformation of the following form.
\begin{equation}
\CF_{\rm E,\al_s}^{\Sigma}\;=\;\int_{\BS}d\mu(\al_t)\;
F_{\rm\sst E}^{\rm\sst L}(\al_s\,|\,\al_t)\;\CG_{\rm E,\al_t}^{\Sigma}\; .
\end{equation}
The kernel $F_{\rm\sst E}^{\rm \sst L}(\al_s\,|\,\al_t)$ is given by the 
following expression:
\begin{equation}\label{fuscoeff1}\begin{aligned}
F_{\rm\sst E}^{\rm \sst L}(\,\al_s\,|\,\al_t)\;=\;
\frac{s_b(u_1)}{s_b(u_2)}\frac{s_b(w_1)}{s_b(w_2)}
\int\limits_{\BR}dt\;\,\prod_{i=1}^4\;\frac{s_b(t-r_i)}{s_b(t-s_i)},
\end{aligned}\end{equation}
where the special function $s_b(x)$ 
is related to $e_b(x)$ via
$s_b(x)=
e^{-\frac{\pi i}{2}x^2}e^{\frac{\pi i}{24}(2-Q^2)}e_b(x)$.
The coefficients $r_i$, $s_i$, $u_i$ and $w_i$ are 
\begin{equation}  
\begin{aligned} r_\1=& p_2-p_1,\\
         r_\2=& p_2+p_1, \\
         r_3=& -p_4-p_3,\\
        r_4=& -p_4+p_3,
\end{aligned}\qquad\quad
\begin{aligned} 
       s_\1=& c_b-p_4+p_2-p_t,\\
         s_\2=& c_b-p_4+p_2+p_t,\\
         s_3=& c_b+p_s,\\
        s_4=& c_b-p_s . 
\end{aligned}
\qquad\quad \begin{aligned} u_\1=& p_s+p_2-p_1,\\
         u_\2=& p_s+p_3+p_4,\\
         w_1=& p_t+p_1+p_4,\\
        w_2=& p_t+p_2-p_3 , 
\end{aligned} \end{equation}
where $c_b=i\frac{Q}{2}$ and $\al_\flat=\fr{Q}{2}+ip_{\flat}$
for $\flat\in\{1,2,3,4,s,t\}$.
Setting $\al_t=\fr{Q}{2}+ip_t$
one may finally write the measure $d\mu(\al_t)$ in the form 
$d\mu(\al_t)=dp_t m(p_t)$, where 
$m(p_t)=4\sinh2\pi bp_t\sinh2\pi b^{-1}p_t$. 
\\[1ex]
\noindent
{\sc B-Move}: The B-move is realized simply by the multiplication 
with the phase factor
\begin{equation}
B^{\rm\sst L}(\al_3,\al_2,\al_1)\;=\;e^{\pi i(\De_{\al_3}-\De_{\al_2}-
\De_{\al_1})},
\end{equation}
where $\De_{\al_k}$, $k=1,2,3$ are the conformal dimensions
$\De_{\al}=\al(Q-\al)$.\\[1ex]
{\sc S-Move}? It is not known yet how to construct a representation 
of the  S-Move on the spaces of Liouville conformal blocks.

\section{Length operators}

We have described how to assign to a Riemann surface $\Sigma$ two
vector spaces $\CH^{\rm L}(\Sigma)$ (for genus 0) 
and $\CH^{\rm T}(\Sigma)$ (for arbitrary genus, but $s>0$), each 
equipped with a (projective) representation of the mapping class
group. According to the conjecture of H. Verlinde the two 
are equivalent. One possible strategy to prove this conjecture 
is to construct a representation of the Moore-Seiberg groupoid 
on the quantized Teichm\"uller spaces, allowing one to compare the
respective representations of its generators. 

\subsection{Length-Twist coordinates}

We find it instructive to discuss the corresponding classical story first.
Regarding the Teich\-m\"uller spaces as a collection
of phase-spaces naturally leads one to look for suitable 
Hamiltonians. A natural choice is furnished by the lengths
of geodesics around simple closed curves, considered as functions
on $\CT(\Sigma)$. It is known that the length-functions 
$l_c$, $l_{c'}$ associated to {\it non-intersecting} curves
$c$ and $c'$ Poisson-commute w.r.t. the Weil-Petersson symplectic form
\cite{Wo}.
 
In the case of a Riemann surface $\Sigma_g^s$
of genus $g$
with $s$ boundary components it is well-known that there will be a 
collection of $3g-3+s$ closed geodesics $c_i$ 
such that cutting $\Sigma_g^s$ along $c_i$, $i=1,\dots,3g-3+s$
decomposes it into a collection of pants (three-holed spheres). 
The collection of  
lengths $l_i$ of the geodesics $c_i$ furnishes a set of functions
on the Teichm\"uller space $T(\Sigma_g^s)$ that Poisson-commute 
with each other, $\{l_i,l_j\}=0$. Having half as many commuting
``Hamiltonians'' as ${\rm dim}(\CT(\Sigma_g^s))=6g-6+2s$ we 
thereby recognize the Teichm\"uller space as a completely 
integrable system. 

A nice feature of the Fock coordinates is that they 
lead to a particularly simple way to 
express the lengths $l_\gamma$ of simple closed curves $\gamma$
in terms the variables $z_e$.
Assume given a 
closed path $\CP_\gamma$ on the fat graph homotopic to a
simple closed curve $\gamma$. 
Let the edges $e_i$, $i=1,\dots,r$ be labeled 
according to the order in which they appear on the path $\CP_\gamma$, 
and define $\sigma_i$ to be $1$ if the path turns left\footnote{w.r.t.
to the orientation induced by the imbedding of the fat-graph 
into the surface} 
at the vertex
that connects edges $e_i$ and $e_{i+1}$, and to be equal to $-1$ 
otherwise. The length $l_\gamma $ is then computed as follows \cite{Fo}.  
\begin{equation}\label{fuchsgen}
2\cosh\big(\fr{1}{2}l_{\gamma}\big)\;=\;|{\rm tr}({\rm X}_{\gamma})|\;,
\quad\qquad
{\rm X}({\gamma})\;=\;{\rm V}^{\sigma_r}{\rm E}(z_{e_r})
\dots {\rm V}^{\sigma_1}
{\rm E}(z_{e_1}),
\end{equation}
where the matrices ${\rm E}(z)$ and ${\rm V}$ are defined respectively by
\begin{equation}
{\rm E}(z)\;=\;\left(\begin{array}{cc} 0 & +e^{+\frac{z}{2}}\\
-e^{-\frac{z}{2}} & 0 \end{array}\right),\qquad
{\rm V}\;=\;\left(\begin{array}{rr} 1 & 1 \\ -1 & 0 \end{array}\right).
\end{equation}

Having chosen the $\{l_1,\dots,l_{3g-3+s}\}$ as the set of 
``action''-variables, it is amusing to note that the corresponding 
``angle''-variables are nothing but the twist-{\it angles} corresponding 
to the deformation of cutting $\Sigma_g^s$ along $c_i$ and 
twisting by some angle $\varphi_i$ before gluing back along $c_i$ \cite{Wo}.
The set
$\{l_1,\dots,l_{3g-3+s};\varphi_1,\ldots\varphi_{3g-3+s}\}$
gives another set of coordinates on the Teichm\"uller space $\CT(\Sigma_g^s)$,
the classical Fenchel-Nielsen coordinates. 

Dehn twists along one of the curves $c_i$, $i=1,\dots,3g-3+s$ 
do not change the pants decomposition and 
are described by $\varphi_i\rightarrow\varphi_i+2\pi$. 
It seems natural to associate length-twist coordinates
$\{l_1,\dots,l_{3g-3+s};\varphi_1,\ldots\varphi_{3g-3+s}\}$ to 
each Moore-Seiberg graph $\Ga$. Changing the Moore-Seiberg
graph $\Gamma$ by the Dehn-twist along one of the
curves $c_i$ corresponds to a change of the twist coordinate
$\varphi_i\ra\varphi_i'=\varphi_i+2\pi$. The Moore-Seiberg groupoid 
thereby gets the interpretation as the natural groupoid of 
canonical transformations of the length-twist coordinates.

\subsection{The length operators}  

Our aim is to make contact with the Liouville conformal field theory.
In the Moore-Seiberg formalism for conformal field theories one 
considers bases for the space of conformal blocks that are associated
to pants decompositions of Riemann surfaces. Our task may be seen
as a quantum counterpart of the task to construct the
change of variables from the Penner- to the Fenchel-Nielsen
coordinates.

Let us consider a simple closed curve $c$ in $\Sigma$ 
such that the triangulation in an annular neighborhood of $c$
looks as depicted in Figure \ref{trann}.
In this case one finds according to (\ref{fuchsgen}) an expression for the
hyperbolic cosine of the geodesic length function that is easy to quantize,
\begin{equation}\label{lodef}
\SL_{c}\;=\;2\cosh2\pi b\spp+e^{2\pi b \sx},\qquad
\begin{aligned}
2\pi b \sx \;=&\;\fr{1}{2}(\sz_b-\sz_a), \\
2\pi b \spp\;=&\;\fr{1}{2}(\sz_a+\sz_b).
\end{aligned}
\end{equation}
We shall work in the
representation where $\spp$ is diagonal. It is easy to check that the
functions
\begin{equation}\label{eigenf}
\Psi(s\,|\,p)\;=\;s_b^{-1}(p-s-c_b)s_b^{-1}(p+s-c_b)
\end{equation}
are eigenfunctions of $\SL_{c}$ 
with eigenvalue being $2\cosh 2\pi b s$.
Let $\bra\,{\rm d}_p\,|$ be the distribution represented by
$\Psi(s\,|\,p+i0)$. The following important
result was proven by R. Kashaev
in \cite{Ka4}.
 
\begin{thm} \label{kathm}\cite{Ka4}
The following set of distributions
\[ 
{\mathfrak F}_{\rm\sst E}^{\flat}\;\equiv\;
\big\{\,|\,{\rm d}_p\,\ket \;;
\;p_s\in\BR^+\,\big\},
\]
forms a basis for 
$L^2(\BR)$
in the sense of generalized functions. We have the relations
\begin{equation} \label{orthcomp} \begin{aligned}
{} & \bra\,{\rm d}_p\,|\,{\rm d}_q\,\ket = \;{\rm m}^{-1}(p)\,\de(p-q)
\qquad\qquad\;\text{\rm (Orthogonality)}, \\
{}& 
\int_{\BR^+}dp \;{\rm m}(p) \;|\,{\rm d}_p\,\ket \bra\,{\rm d}_p\,|
\;= \;\id \qquad \qquad\text{\rm (Completeness)},
\end{aligned}\end{equation} 
where the measure $m(p)$ is defined as 
$m(p)=4\sinh 2\pi bp\,\sinh 2\pi b^{-1}p$.
\end{thm}

It follows that there exists a self-adjoint
operator $\sll_{c}$ with spectrum $\BR_+$ 
such that $\SL_{c}=2\cosh\frac{1}{2}\sll_{c}$. 
$\sll_{c}$ is the quantum operator corresponding to the 
hyperbolic length around the geodesic $c$.

It will be shown \cite{TT} that the definition of the length operators 
$\sll_{\ga}$ can be extended to {\it arbitrary} closed geodesics $\ga$
such that 
(i) $\sll_{\ga}$ is self-adjoint with spectrum $\BR_+$, and
(ii) $[\sll_{\ga},\sll_{\ga'}]=0$ if $\ga\cap\ga'=\emptyset$.
Moreover, diagonalization of the length operator  
$\sll_{\ga}$ for a closed geodesic $\ga\subset\Sigma$ leads to a 
{\it factorization} of $\pi(\Sigma,\Lambda)$ 
in the following sense. Let $\Sigma'_{\ga}\equiv
\Sigma\setminus\ga$ be the possibly
disconnected Riemann surface obtained by cutting along $\ga$.
The coloring $\Lambda$ of the boundary components of $\Sigma$ can be
naturally extended to a coloring 
$\Lambda'_{\ga,l}$ for $\Sigma'_{\ga}$ by assigning the number $l\in\BR^+$ to
the two new boundary components that were created by
cutting along $\ga$. The spectral representation for $\sll_\ga$ then
yields the following representation for $\pi(\Sigma,\Lambda)$.
\begin{equation}
\pi(\Sigma,\Lambda)\;\simeq\; \int_{\BR^+}^{\oplus}dl \;
\pi(\Sigma'_{\ga},\Lambda'_{\ga,l}).
\end{equation}
The corresponding representations of the mapping class group 
factorize/restrict accordingly \cite{TT}.
This allows one to construct bases (in the sense of generalized functions)
for $\CH(\Sigma)$ labeled by the assignments of lengths to the 
closed geodesics $c_1,\dots,c_{3g-3+s}$ that define a pants 
decomposition. 

\section{Realization of the Moore-Seiberg groupoid on $\CH(\Sigma)$}

Thanks to the factorization properties of the quantized Teichm\"uller
spaces one may indeed construct 
a realization of the Moore-Seiberg groupoid
by associating unitary operators to the elementary moves depicted in
Figures  \ref{MS1}-\ref{MS3} \cite{TT}. 

\subsection{A-Move} 
In order to construct the representation of the
A-move let us consider the four-holed sphere with the fat graph 
introduced in Figure \ref{fourpt}.
The operators $\SL_s$ and $\SL_t$ 
that represent the lengths of the geodesics isotopic 
to the s- and t-cycles drawn in Figure \ref{fourpt} respectively
are then given by the following expressions. 
\begin{equation}\begin{aligned}
\SL_s\;=\;& 2\cosh2\pi b \spp_s\\
& +e^{\pi b\sx}\;
2\cosh \pi b(\spp_s+\mss_1-\mss_2) 2\cosh \pi b(\spp_s+\mss_3+\mss_4)\;
 e^{\pi b\sx}\;, \\
\SL_t\;=\;& 2\cosh2\pi b \spp_t\\
& +e^{-\pi b\sx}\;
2\cosh \pi b(\spp_t+\mss_2-\mss_3) 2\cosh \pi b(\spp_t-\mss_4-\mss_1)\;
 e^{-\pi b\sx} \;,
\end{aligned}
\end{equation}
where we have used the notation $\spp_s=\spp-\mss_3-\mss_4$,
$\spp_t=\spp-\mss_3-\mss_2$ with
\begin{equation}
\spp=(2\pi b)^{-1}\sz_5,\qquad
\sx=(2\pi b)^{-1}\sz_1,\qquad \mss_k=(4\pi b)^{-1}\sfc_k, \quad
k=1,\dots,4.
\end{equation} 
In order to find eigenfunctions of the operators $\SL_s$ and $\SL_t$
let us observe that the ansatz
\begin{equation}\label{eigenans}
|\,{\rm d}^{\flat}_p\,\ket\;=\;\rho_{\sst{\rm D},p}^{\flat}\,
\SU_{\rm\sst D}^{\flat}
|\,{\rm d}_p\,\ket,\qquad
\flat=s,t,\quad p\in\BR^+.
\end{equation}
allows one to construct eigenstates of $\SL_s$ and $\SL_t$ from the 
eigenstates $|\,{\rm d}_p\,\ket$ 
of the operator $\SL_c$ that were defined in (\ref{eigenf}),
provided that
the unitary operators $\SU_{\rm\sst D}^{\flat}$ are chosen as
\begin{equation}\label{efdef}
\begin{aligned}
\SU_{\rm\sst D}^{s}\;=\;&e^{2\pi i (s_3+s_4)\sx}
s_b(\spp+s_1-s_2)s_b(\spp+s_3+s_4),\\
\SU_{\rm\sst D}^{t}\;=\;&e^{2\pi i (s_2+s_3)\sx}
     s_b^{-1}(\spp+s_2-s_3)s_b^{-1}(\spp-s_1-s_4)\;.
\end{aligned}
\end{equation}
The tuple ${\rm D}=(s_4,\dots,s_1)$ parameterizes 
the fixed lengths 
of the four boundary holes via $l_k=4\pi bs_k$, $k=1,\dots,4$,
and the $\rho_{\sst{\rm D},p}^{\flat}$ are pure phases that will be fixed 
shortly.  
We have
\begin{thm} \label{complthm} \cite{PT}
The following two sets of distributions
\[ 
{\mathfrak F}_{\rm\sst E}^{\flat}\;\equiv\;
\big\{\,|\,{\rm d}_p^\flat\,\ket \;;
\;p\in\BR^+\,\big\}, \qquad
\flat\in\{s,t\}
\]
form bases for 
$L^2(\BR)$
in the sense of generalized functions. We have the relations
\begin{equation} \label{orthcomp2} \begin{aligned}
{} & \bra\,{\rm d}_p^\flat\,|\,{\rm d}_q^\flat\,\ket = \;{\rm m}^{-1}(p)\,\de(p-q)
\qquad\qquad\;\text{\rm (Orthogonality)}, \\
{}& 
\int_{\BR^+}dp \;{\rm m}(p) \;|\,{\rm d}_p^\flat\,\ket \bra\,{\rm d}_p^\flat\,|
\;= \;\id \qquad \qquad\text{\rm (Completeness)}.
\end{aligned}\end{equation} 
\end{thm} 
This theorem was first obtained by another method \cite{PT}, 
but thanks to (\ref{eigenans}) it also follows easily from
Kashaev's Theorem \ref{kathm}. Another simple implication
of Theorem \ref{complthm} is the relation
\begin{equation}
\int_{\BR^+}dq\,{\rm m}(q)\;
|\,{\rm d}_{q}^t\,\ket\;F_{\rm\sst E}^{\rm\sst T}(\,p\,|\,q\,)\;
=\; |\,{\rm d}_{p}^s\,\ket,\qquad
F_{\rm\sst E}^{\rm\sst T}(\,p\,|\,q\,)\;\equiv\; 
\bra\,{\rm d}_q^t\,|\,{\rm d}_p^s\,\ket\; .
\end{equation}

The unitary operator that represents the A-move will simply be the
one that describes the change of basis between 
the eigenfunctions of the length operators $\sll_s$ and $\sll_t$ respectively.
It is represented 
by the kernel
$F_{\rm\sst D}^{\rm\sst T}(\,p\,|\,q\,) =
\bra \,{\rm d}^{t}_q\,|\,{\rm d}^{s}_p\,\ket$.
It is natural to choose the phase factors $\rho_{\sst{\rm D},p}^{\flat}$
in such a way that the b-6-j symbols $\{\dots\}_b^{}$ defined by 
\begin{equation}
F_{\rm\sst D}^{\rm\sst T}(\,q\,|\,p\,)\;=\;\big\{\begin{smallmatrix}
s_1 & s_2 & p \\ s_3 & s_4 & q \end{smallmatrix}\big\}
\end{equation}
satisfy the following tetrahedral symmetries
\begin{equation}\label{tetr}
\big\{\begin{smallmatrix}
s_1 & s_2 & s_3 \\ s_4 & s_5 & s_6 \end{smallmatrix}\big\}_b^{}
=\big\{\begin{smallmatrix}
s_4 & -s_5 & -s_3 \\ s_1 & -s_2 & -s_6 \end{smallmatrix}\big\}_b^{}
=\big\{\begin{smallmatrix}\phantom{-}
s_2 & \phantom{-}s_1 & \phantom{-}s_3 \\ 
-s_5 & -s_4 & -s_6 \end{smallmatrix}\big\}_b^{}
=\big\{\begin{smallmatrix}
\phantom{-}s_1 & -s_3 & -s_2 \\ 
-s_4 & -s_6 & -s_5 \end{smallmatrix}\big\}_b^{}\;\, .
\end{equation}
Let us choose the $\rho_{\sst{\rm D},p}^{\flat}$ as follows:
\begin{equation}
\rho_{\sst{\rm D},p}^{s}\;=\;\frac{s_b(p+s_2-s_1)}{s_b(p+s_3+s_4)},\qquad
\rho_{\sst{\rm D},q}^{t}\;=\;\frac{s_b(q+s_2-s_3)}{s_b(p+s_1+s_4)}\;.
\end{equation}
The kernel $F_{\rm\sst D}^{\rm\sst T}(\,q\,|\,p\,)$ 
will then differ from 
the b-Racah-Wigner coefficients 
for the quantum group $\CU_q(\fsl(2,\BR))$ studied and calculated 
in \cite{PT} only by multiplication with the measure ${\rm m}(q)$.
It follows from the results of \cite{PT3} (Appendix B.2) that 
the corresponding 
b-6-j symbols indeed satisfy (\ref{tetr}).
\subsection{B-move}

\begin{figure}[h]
\begin{center}\epsfxsize6cm
\epsfbox{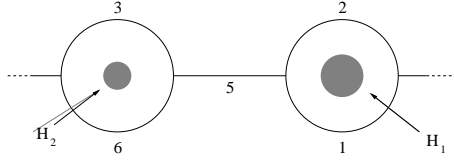}
\end{center}
\caption{Fat graph in the neighborhood of the two relevant holes.}
\label{bfat}
\end{figure}

In order to discuss the B-move we find it convenient to use the Fock-variables
associated to the edges of the fat graph depicted in Figure \ref{bfat}. 
Let us furthermore introduce the following linear combinations of the
variables $\sz_e$, $e\in\{1,2,3,5,6\}$.
\begin{equation}
\begin{aligned}
\spp_2=&\,(4\pi b)^{-1}(\sz_6-\sz_3),\\
\spp_1=&\,(4\pi b)^{-1}(\sz_1-\sz_2),
\end{aligned}\qquad
\begin{aligned}
\mss_2=&\,(4\pi b)^{-1}(\sz_6+\sz_3),\\
\mss_1=&\,(4\pi b)^{-1}(\sz_1+\sz_2),
\end{aligned}\qquad
\begin{aligned}
\sq_2=&\,-(2\pi b)^{-1}\sz_5,\\
\sq_1=&\,+(2\pi b)^{-1}\sz_5.
\end{aligned}
\end{equation}
The representation of the braiding of two punctures in terms 
of the generators of the Ptolemy groupoid was first discussed in
\cite{Ka3}. Considering punctures (holes of zero size) corresponds 
to setting $\mss_2=0=\mss_1$. The expression for the 
generator of the B-move that was found in 
\cite{Ka3} may be written as 
\begin{equation}
\SR\;=\;e^{\pi i\spp_2\spp_1}\,\frac{s_b(\spp_1)}{s_b(\spp_2)}\,
e_b^{-1}(\sq_1-\sq_2)\,
\frac{s_b(\spp_2)}{s_b(\spp_1)}\,e^{\pi i\spp_2\spp_1}\; .
\end{equation}
As already observed in \cite{Ka3}, $\SR$ is identical to the 
R-operator for the modular double $\CD\CU_q(\fsl(2,\BR))$ of
$\CU_q(\fsl(2,\BR))$ that was first proposed in \cite{Fa}
and further studied in \cite{BT}. By working out the operator
$\SL_{21}$ that represents the length of the geodesic 
surrounding the two punctures via $\SL_{21}=2\cosh\frac{1}{2}\sll_{21}$,
one finds that $\SL_{21}$ coincides with the realization of the
Casimir $\SC_{21}$ of $\CD\CU_q(\fsl(2,\BR))$ in the
tensor product of two of its representations.
 
These observations may then be used in order to find the operator
$\SR_{s_2s_1}^{}$ that represents the braiding of two 
holes with 
finite circumferences $l_k=(4\pi b)^{-1}s_k$. In order to do this
one may ``fill'' the holes by gluing disks with two punctures each
to the boundaries of the holes. The braiding of the two holes can be
decomposed into a sequence of braidings of the four punctures. By using
\cite{BT} (Theorem 3) it is then possible to show that
\begin{equation}
\SR_{s_2s_1}^{}\;=\;e^{\pi i\spp_2\spp_1}\,
\frac{s_b(\spp_1-\mss_1)}{s_b(\spp_2+\mss_2)}\,
e_b^{-1}(\sq_1-\sq_2)\,
\frac{s_b(\spp_2+\mss_2)}{s_b(\spp_1-\mss_1)}\,e^{\pi i\spp_2\spp_1}\; .
\end{equation}
It remains to observe that according to \cite{BT} (Theorem 6)
the operator $\SR_{s_2s_1}^{}$ becomes diagonal on eigenstates of the
length operator $\SL_{21}^{}$. The eigenvalue of $\SR_{s_2s_1}^{}$ is
\begin{equation}
B^{\rm T}(l_3,l_2,l_1)\;=\;
e^{\pi i(\Delta_{l_3}-\Delta_{l_2}-\Delta_{l_1})},
\qquad \Delta_l=\frac{Q^2}{4}
+\Bigl(\frac{l}{4\pi b}\Bigr)^2.
\end{equation}

\subsection{S-move} One possibility to introduce the
operator that represents the S-move is to consider the 
change of basis between bases that diagonalize the 
length operators of a- and b- cycles on the once-holed torus respectively.
The proof that the operators that represent the A-, B- and
S-moves satisfy the relations of the Moore-Seiberg 
groupoid \cite{MS1,BK} will be given in
\cite{TT}. Concerning the relations coming from genus zero surfaces
we may observe that their validity immediately follows if one
combines the above observations concerning relations with 
$\CD\CU_q(\fsl(2,\BR))$ with the results of \cite{PT,BT}.

It furthermore turns out that the kernel that represents
the S-move can be expressed in terms of $(F^{\rm T},B^{\rm T})$ 
by means of an expression which is similar to the
one that was found for rational conformal field theories in \cite{MS2}.
The data $(F^{\rm T},B^{\rm T})$ are therefore sufficient
to characterize the projective representation
of the Moore-Seiberg groupoid on the quantized Teichm\"uller
spaces $\CH_g^s$ completely.
 
\subsection{The Verlinde conjecture}

The main observation to be made is the following. One has
\begin{equation}\label{verl}
(F^{\rm T},B^{\rm T})\;\equiv\; (F^{\rm L},B^{\rm L})
\quad \text{provided that}\quad \al_k=\frac{Q}{2}+i\frac{l_k}{4\pi b}.
\end{equation}
It follows that the spaces of conformal blocks of Liouville theory  
and the Hilbert spaces from the quantization of the Teichm\"uller spaces
are indeed isomorphic as representations of the mapping class group.

\section{Final remarks}

First let us note that
so far the conformal blocks of Liouville theory were only constructed
in genus zero, but our results imply that the 
corresponding mapping class group representation has a 
consistent and essentially unique extension to higher genus. 

One may wonder if there is a more direct relation between the conformal 
blocks of Liouville theory as discussed in Section 4 and the eigenfunctions
of the length operators from Sections 5 and 6. 
In this regard one may observe that
so far the quantization of the Teichm\"uller spaces $\CT(\Sigma)$
was based on the choice of a real polarization for $\CT(\Sigma)$. 
However, the complex structure on the Teichm\"uller spaces $\CT(\Sigma)$
should allow one to introduce an alternative (``coherent state'') 
representation where the states are represented by wave-functions that
are analytic on $\CT(\Sigma)$. Such a representation should have the 
property that the action of the mapping class group (the symmetry group
of our phase-space) is represented by monodromy transformations. The
field theoretical conformal blocks have exactly this property. 
This leads us to suggest that the Liouville conformal blocks are in
fact just representing the change of basis from a coherent
state basis for $\CH(\Sigma)$ to a basis in which the length operators
are diagonal.

\begin{figure}[t]
\epsfxsize8cm
\centerline{\epsfbox{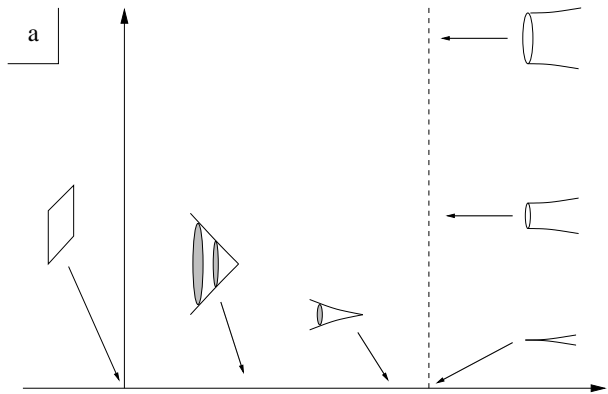}}
\caption{Relation between the parameter $a$ and the ``shape'' of the 
boundary.}\label{alfig}
\end{figure}

Our results finally establish a precise correspondence (\ref{verl})
between
the labels $\al_k=\frac{Q}{2}+iP$ of a Liouville vertex operator $\SV_{\al}$
and the length $l_k$ of the hole to which $\SV_{\al}$ is associated.
The fact that all the basic objects are meromorphic w.r.t.
the variables $\al_{k}$ or $l_{k}$ allows one to analytically continue 
from the case considered above to cases where $\al_{k}\in\BR$.
The geometrical interpretation of the cases $\al_{k}\in\BR$ was 
previously inferred from the semi-classical 
relation between $\al$ and the type
of singularity that the metric $ds=e^{2\varphi}dzd\bar{z}$ 
develops near the insertion point of $\SV_{\al}$, see e.g. \cite{Se}. 
Combining these
two observations gives the diagram drawn in Figure \ref{alfig},
which relates the parameter $\al$ to the ``shape'' of the boundary 
component to which $\SV_{\al}$ is associated. It is remarkable and
nontrivial that the point $\al=0$ indeed corresponds to the complete 
disappearance of a boundary component, as follows from 
\cite{PT3} (Appendix B.1).

\bigskip
\noindent{\bf Acknowledgments} $\frac{\quad}{}$
The author would like to thank the organizers of the 6th 
International Conference on CFTs and
Integrable Models, Chernogolovka, for organizing a nice and
stimulating conference. Many thanks also to L. Checkov, V. Fock
and especially R. Kashaev for discussions and explanations of their
work.

I am furthermore grateful to the SPhT CEA-Saclay,
where this work was completed, for hospitality. 
I would finally like to acknowledge support by the SFB 288 of the DFG.



\begin{thebibliography}{77}
\newcommand{\CMP}[3]{{Comm. Math. Phys. }{\bf #1} (#2) #3}
\newcommand{\LMP}[3]{{Lett. Math. Phys. }{\bf #1} (#2) #3}
\newcommand{\PL}[3]{{Phys. Lett. B}{\bf #1} (#2) #3}
\newcommand{\NP}[3]{{Nucl. Phys. B}{\bf #1} (#2) #3}
\bibitem{Pol} A.M. Polyakov: 
   \PL{103}{1981}{207-210}
\bibitem{Kr} K. Krasnov, 
   Class.Quant.Grav. {\bf 19} (2002) 3977-3998 [arXiv: hep-th/0112164]
\bibitem{KV} D. Klemm, L. Vanzo,  
   JHEP {\bf 0204} (2002) 030 [arXiv: hep-th/0203268]
\bibitem{ZT} P.G. Zograf, L. Takhtajan, 
   Math. USSR Sb. {\bf 60} (1988) 297-313
\bibitem{TL} J. Teschner:  
  {Class.\ Quant.\ Grav.} {\bf 18} (2001)  R153-R222
  [arXiv: hep-th/0104158]
\bibitem{TL2} J. Teschner: A lecture on the Liouville vertex operators,
   in this volume.
\bibitem{V} H. Verlinde, 
   \NP{337}{1990}{652-680}
\bibitem{Fo} V.V. Fock,
   [arXiv: hep-th/9702018] 
\bibitem{Ka1} R.M. Kashaev, 
   \LMP{43}{1998}{105-115}, [arXiv: q-alg/9705021]   
\bibitem{CF} L. O. Checkov, V. Fock, 
  Theor. Math. Phys. {\bf 120} (1999) 1245-1259\\
  {[arXiv: math.QA/9908165]}
\bibitem{Pe} R.C. Penner: 
   \CMP{113}{1987}{299-339}
\bibitem{IT} Y. Imayoshi, M. Taniguchi: {\it An introduction to 
    Teichm\"uller spaces}, Springer
\bibitem{Ge} S. Gervais, 
   {Trans. AMS}{\bf 348} (1996) 3097-3132
\bibitem{BPZ} A.A. Belavin, A.M. Polyakov, A.B. Zamolodchikov,
  \NP{241}{1984}{333}
\bibitem{MS1} G. Moore, N. Seiberg, 
   \CMP{123}{1989}{177-254}
\bibitem{BK} B. Bakalov, A. Kirillov jun.,
  Transform. Groups {\bf 5} (2000) 207-244\\ 
  {[arXiv: math.GT/9809057]}
\bibitem{Ka2} R.M. Kashaev,
   Proc. Steklov Inst. of Math. {\bf 226} (1999) 63-71, \\
   {} [arXiv: hep-th/9811203]
\bibitem{Ka3} R.M. Kashaev, 
  [arXiv: math.QA/0008148]
\bibitem{Ka4} R.M. Kashaev, 
  In: S. Pakuliak and G. von Gehlen (eds.), Integrable structures of 
  exactly solvable two-dimensional models of quantum field theory 
  (Kiev, 2000), 211--221, NATO Sci. Ser. II Math. Phys. Chem., 35, 
  Kluwer Acad. Publ., Dordrecht, 2001.
\bibitem{TT} J. Teschner, paper in preparation.
\bibitem{Wo} S.A. Wolpert, 
   Amer. J. Math {\bf 107} (1985) {969-997}
\bibitem{PT} B.Ponsot, J.Teschner, 
    \CMP{224}{2001}{613-655}\\{} [arXiv: math.QA/0007097]
\bibitem{PT3} B.Ponsot, J.Teschner, 
  \NP{622}{2002}{309-327}
\bibitem{Fa} L.D. Faddeev, 
    Math. Phys. Stud. {\bf 21} (2000) 149-156  [arXiv: math.QA/9912078]
\bibitem{BT} A. Bytsko, J. Teschner,
  [arXiv: math.QA/0208191]
\bibitem{MS2} G. Moore, N. Seiberg, 
   in: Proceedings of the Trieste Spring School 1989, p. 1-129 
\bibitem{Se} N. Seiberg: 
  Progr. Theor. Phys. Suppl. {\bf 102} (1990) {319}
\end{thebibliography}
\end{document}